\documentclass[sigconf]{acmart}

\AtBeginDocument{%
  }

\copyrightyear{2026}
\acmYear{2026}
\setcopyright{cc}
\setcctype{by}
\acmConference[IWSPA '26]{Proceedings of the 12th ACM International Workshop on Security and Privacy Analytics}{June 23--25, 2026}{Frankfurt am Main, Germany}
\acmBooktitle{Proceedings of the 12th ACM International Workshop on Security and Privacy Analytics (IWSPA '26), June 23--25, 2026, Frankfurt am Main, Germany}
\acmDOI{10.1145/3806007.3810966}
\acmISBN{979-8-4007-2609-5/2026/06}

\usepackage{amsmath}

\usepackage{amssymb,amsfonts}
\usepackage{graphicx}
\usepackage{booktabs}
\usepackage[normalem]{ulem}
\usepackage{xcolor}
\usepackage{balance}

\begin{document}

\title{Robust Ensemble of Selectively Strengthened and Augmented Predictors}

\author{Parsa Memarzadehsaghezi}
\authornote{These authors contributed equally to this work.}
\affiliation{%
  \institution{Ontario Tech University}
  \city{Oshawa}
  \state{Ontario}
  \country{Canada}
}
\email{parsa.memarzadeh@ontariotechu.ca}

\author{Zahra Hashemi}
\authornotemark[1]
\affiliation{%
  \institution{Ontario Tech University}
  \city{Oshawa}
  \state{Ontario}
  \country{Canada}
}
\email{zahra.hashemi@ontariotechu.ca}

\author{Pooria Madani}
\affiliation{%
  \institution{Ontario Tech University}
  \city{Oshawa}
  \state{Ontario}
  \country{Canada}
}
\email{pooria.madani@ontariotechu.ca}

\author{Mehran Ebrahimi}
\affiliation{%
  \institution{Ontario Tech University}
  \city{Oshawa}
  \state{Ontario}
  \country{Canada}
}
\email{Mehran.Ebrahimi@ontariotechu.ca}

\begin{abstract}
Evasion attacks present a significant challenge to the robustness of machine learning (ML)-based classifiers, particularly in critical applications such as fraud detection and cybersecurity. Although existing defense mechanisms are effective in some settings, they often suffer from limited generalizability and do not systematically improve model robustness across diverse attack scenarios. To address these limitations, we introduce Robust Ensemble of Selectively Strengthened and Augmented Predictors (\textit{RESSAP}), a novel framework that transforms a single classifier into an ensemble of robust classifiers. Each classifier in the ensemble is trained on a carefully selected subset of features, where feature selection is guided by a resilience metric that accounts for both feature importance and robustness. During inference, a \emph{random} subset of these classifiers is used to make predictions, increasing unpredictability and improving resistance to adversarial manipulation. In addition, noise-based data augmentation is applied during training to strengthen decision boundaries and improve generalization. Our experimental results demonstrate that RESSAP significantly improves robustness against adversarial evasion attacks while maintaining strong accuracy on clean data. Overall, this model-agnostic framework provides a scalable and flexible defense strategy for enhancing the security of machine learning systems without requiring major changes to existing architectures.
\end{abstract}



\maketitle
\section{Introduction}
Despite the remarkable success of machine learning (ML) in classification tasks, these models often exhibit blind spots---regions in the input space where their predictions become unreliable. Such blind spots arise naturally due to the limited scope of training data, the high dimensionality of feature spaces, and the inherent approximations made by learning algorithms. As a result, even well-performing models may misclassify inputs that fall just outside familiar patterns, leaving them vulnerable to deliberate manipulation in adversarial settings, commonly known as adversarial evasion attacks~\cite{biggio2018wild, goodfellow2015explaining, madry2018towards}. These attacks exploit blind spots in learned decision boundaries, causing input instances to be misclassified at runtime without affecting the training process.

For example, in financial fraud detection systems that rely on anomaly detection models, an adversary may split a large transaction into several smaller ones, each carefully crafted to remain below the model’s learned detection thresholds. In doing so, the attacker can mimic legitimate spending behaviour and evade detection by exploiting the model’s limited exposure to such subtle manipulations during training. Such evasion attacks are especially concerning because they exploit the same generalization mechanisms that learning algorithms rely on to form decision boundaries from training data, making them difficult to detect and potentially costly in high-stakes applications~\cite{sharif2016accessorize}.

In adversarial settings, attackers often have the ability to interact with a deployed ML model either through direct query access or by observing its outputs. This interaction enables them to probe the model’s behaviour and iteratively refine their inputs in order to craft instances (i.e., adversarial examples) that evade detection or cause misclassification~\cite{nelson2010near, lowd2005reverse}. Even without access to the internal details of the classification model, repeated interaction with the system can reveal enough information about its decision boundaries to construct highly effective adversarial examples~\cite{chen2020rays, shi2023cisa}.

Despite extensive research on defending against adversarial examples, there remains a lack of \textit{systematic frameworks} for evaluating and improving the robustness of learning-based detection systems against evasion attacks~\cite{biggio2011bagging, rubinstein2009antidote}. Many existing defenses are evaluated under narrowly defined attack scenarios, which limits their generalizability and practical applicability. In addition, many proposed methods require substantial changes to the training process or restrict the choice of learning algorithms, making them impractical for many data scientists and domain experts~\cite{madry2018towards, zhang2015feature}. Consequently, there is a strong need for a more systematic, model-agnostic defense strategy that can improve robustness without compromising flexibility, particularly within modern \textit{Sec-DevOps} pipelines, where the security and robustness of learning-based assets (e.g., models) must be continuously evaluated and improved.

Among the defense strategies explored in the literature, introducing randomness at inference time (e.g., stochastically flipping the classification label) has emerged as a promising approach for hindering adversarial evasion attacks~\cite{cohen2019certified, pinot2020randomization, madanirandomization, madani2022randomized}. Randomization can be applied to input data, model parameters, or decision outputs to reduce an attacker’s ability to exploit consistent model behaviour. By making ML model responses less predictable, these techniques increase the difficulty of reliably crafting successful evasion attempts. However, if applied naively, such randomness can also degrade the classification accuracy of benign inputs. Therefore, any randomized defense must carefully balance robustness and classification accuracy in order to remain practically effective~\cite{pinot2020randomization}.

To address this challenge, we propose Robust Ensemble of Selectively Strengthened and Augmented Predictors (RESSAP), a novel model-agnostic framework that transforms an existing trained ML model into an ensemble of specialized sub-models through feature-space subsetting. At inference time, a random subset of these sub-models is selected to produce predictions, thereby obtaining the benefits of randomization-based defense without compromising overall classification accuracy. Applying this framework within existing ML pipelines significantly enhances the robustness of trained models against adversarial evasion attacks by increasing the number of queries required to craft successful adversarial examples~\cite{cohen2019certified, pinot2020randomization, biggio2011bagging, madanirandomization}. This increased query cost, introduced through randomization, forms the basis of our evaluation, where we demonstrate improved robustness with minimal impact on classification accuracy.

The key contributions of this work are as follows:
\begin{enumerate}
    \item We propose \textbf{RESSAP}, a novel ensemble framework that enhances adversarial robustness by combining feature-level diversity with classifier-level randomization.
    \item We introduce a new feature selection approach that combines permutation importance and robustness into a unified resilience metric for guiding the formation of diverse feature subsets.
    \item We validate the effectiveness of our approach through experiments on synthetic datasets, demonstrating significant improvements over traditional models in both clean accuracy and attack resistance.
\end{enumerate}

The remainder of the paper is organized as follows. Section~\ref{sec:RW} reviews related work on evasion attacks and defense strategies. Section~\ref{sec:threat-model} defines the adversarial threat model. Section~\ref{sec:model} presents our proposed RESSAP framework, including feature selection, data augmentation, ensemble training, and inference. Section~\ref{sec:experiments} describes the experimental setup and presents the results and analysis. Finally, Section~\ref{sec:conclusion} concludes the paper and outlines directions for future work.

\section{Related Work}
\label{sec:RW}

\subsection{Evasion Attacks}
The vulnerability of machine learning models to adversarial manipulation has been extensively studied in recent years~\cite{biggio2018wild}. These studies primarily aim to expose model weaknesses and assess the security risks that arise when models operate in adversarial environments. Among the different attack settings, evasion attacks at test time have received significant attention as a critical research direction.

Goodfellow et al.~\cite{goodfellow2015explaining} introduced one of the earliest gradient-based attacks for neural networks, known as the Fast Gradient Sign Method (FGSM). Their method perturbs inputs in the direction of the loss gradient to generate adversarial examples with minimal computational effort. This simple but effective approach highlighted the sensitivity of deep models to small, targeted perturbations and established the foundation for many subsequent white-box attack methods.

Early work by Nelson et al.~\cite{nelson2010near} introduced one of the first frameworks for computing near-optimal evasion attacks against convex classifiers. Their multi-line search algorithm probes a classifier’s decision boundary along strategically chosen feature directions, combined with binary search, in order to find minimally perturbed inputs that evade detection.

Kantchelian et al.~\cite{kantchelian2016evasion} later showed that even ensemble models such as random forests and boosted trees are highly vulnerable to evasion attacks. They formulated the problem of evading tree ensembles as an optimization task for generating minimal perturbations that can fool all trees in the ensemble. Their results demonstrated that tree ensembles, like linear models, can be effectively compromised by carefully crafted adversarial instances.

Alzantot et al.~\cite{alzantot2019genattack} observed that earlier black-box attacks often required excessive numbers of queries, particularly when building substitute models or estimating gradients. To address this, they introduced GenAttack, a gradient-free black-box attack based on a genetic algorithm that evolves adversarial inputs. Their method achieved competitive success rates on image classifiers while substantially reducing query overhead.

Chen and Gu~\cite{chen2020rays} proposed RayS, a hard-label black-box attack that searches over discrete input directions to locate decision boundaries. RayS removes the need for gradient or confidence information and reduces the number of required queries through a priority-driven ray search strategy.

More recently, Shi et al.~\cite{shi2023cisa} introduced CISA, an adaptive iterative attack that dynamically adjusts sampling and step sizes based on the estimated distance to the classifier’s decision surface. Their method further improves query efficiency while maintaining strong attack performance in limited-feedback black-box settings.

\subsection{Defense Against Evasion Attacks}
Defending classifiers against evasion attacks has been approached from both feature-centric and ensemble-based perspectives. At the feature level, Zhang et al.~\cite{zhang2015feature} proposed an adversary-aware feature selection strategy for training robust classifiers. Their method explicitly optimizes the feature set by trading off nominal accuracy for improved resistance to evasion, producing models that maintain strong performance while becoming more difficult to evade through feature manipulation.

Kołcz and Teo~\cite{kolcz2009feature} introduced a related feature-weighting approach that adjusts the importance of individual features based on robustness. Their method reweights features using the output of an initial classifier and improves resilience in spam detection tasks with relatively low computational cost. Although designed for linear models, their work highlights the value of feature-level adjustments in mitigating adversarial impact.

Another feature-focused line of work considers adversarial example detection rather than direct prevention. Huang et al.~\cite{huang2019perturbations} proposed a model-agnostic detection scheme based on injecting random input perturbations. By analyzing the variance of a model’s outputs under these perturbations, their method can flag inputs that cause abnormal output fluctuations as likely adversarial, without modifying the underlying classifier.

Ensemble methods have also shown promise as a defense strategy. Smutz and Stavrou~\cite{smutz2016tree} leveraged the intrinsic diversity of ensembles to detect evasion attempts. They proposed monitoring the level of agreement among an ensemble’s base classifiers: when an input causes substantial disagreement (i.e., low consensus), the system outputs an ``uncertain'' label rather than a standard prediction. This mutual-agreement analysis forces an attacker to fool multiple diverse models simultaneously, thereby increasing the difficulty of a successful evasion attempt and enabling many attacks to be identified as outliers in the agreement measure.

Pang et al.~\cite{pang2019adp} introduced a more formal approach to ensemble robustness by explicitly promoting prediction diversity. Their Adaptive Diversity Promoting (ADP) regularizer encourages ensemble members to produce distinct non-top-ranked predictions, thereby improving robustness by reducing the transferability of adversarial examples across models. Although effective, their approach primarily emphasizes output-level diversity and does not explore feature-level or model-specific diversity strategies.

In addition to detection and diversity-based methods, randomized smoothing has emerged as a powerful ensemble-inspired defense. Cohen et al.~\cite{cohen2019certified} showed that averaging a base classifier’s predictions over multiple noisy copies of an input, generated using Gaussian noise, produces a ``smoothed'' classifier with certified robustness guarantees. This approach reduces the effect of any single adversarial perturbation, making the overall classifier provably robust up to a quantifiable perturbation radius.

\begin{figure*}[t]
    \centering
    \includegraphics[width=\linewidth]{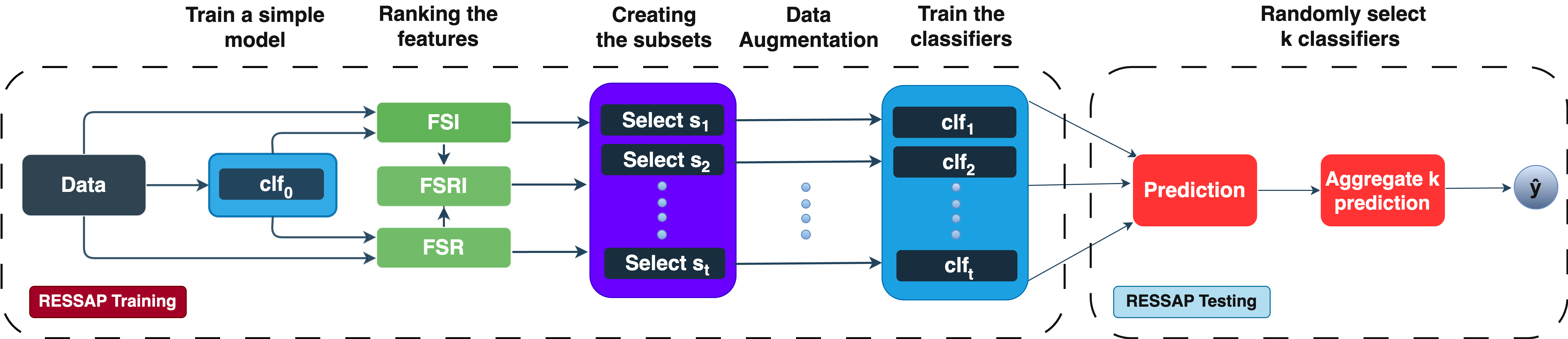}
    \caption{The RESSAP architecture.}
    \Description{High-level diagram of the RESSAP pipeline showing feature scoring, subset construction, augmentation, ensemble training, and randomized inference.}
    \label{fig:architecture}
\end{figure*}

\section{Threat Model}
\label{sec:threat-model}

We consider a black-box evasion setting in which an adversary interacts with a deployed machine learning model by submitting inputs and observing the predicted class labels. Our threat model is based on the following assumptions:
\begin{itemize}
    \item \textbf{Model-Agnostic Deployment:} We assume that an initial machine learning model, denoted as $clf_0$, has already been trained by the defender prior to deployment. Our defense strategy is model-agnostic in the sense that it does not impose restrictions on the architecture or training procedure of $clf_0$. Instead, it operates as a post-training enhancement step, consistent with secure development and operations (SecDevOps) practices. The pretrained model and its associated training data are provided as inputs to our framework, which then applies its defense mechanisms before deployment in adversarial environments.

    \item \textbf{Binary Classification Assumption:} We focus on binary classification tasks (e.g., malicious vs.\ non-malicious), where the negative class represents benign inputs and the positive class represents malicious ones. Nevertheless, without loss of generality, the proposed framework can be extended to multiclass settings and is compatible with a broad range of underlying machine learning models.

    \item \textbf{Adversarial Goal:} The adversary begins with a malicious input $x^A$ that is correctly classified as belonging to the positive (i.e., malicious) class. The adversary’s objective is to find a perturbed instance $x^*$ such that the classifier outputs the negative (i.e., benign) class label.

    \item \textbf{Query Access:} The adversary can submit any syntactically valid input to the deployed model and observe the corresponding output label. Although query access is unrestricted in principle, the adversary seeks to minimize the number of queries because repeated interactions may increase detection risk or operational cost.

    \item \textbf{No Training Control:} The adversary has no influence over the training data, training procedure, or feature selection process. All adversarial manipulations take place at inference time, with the goal of discovering an evading instance $x^*$.

    \item \textbf{Attack Methodology:} We assume that the adversary employs a Multi-Line Search (MLS) strategy constrained by the $\ell_2$ norm~\cite{nelson2010near}. This attack efficiently probes decision boundaries by querying along weighted feature directions. However, the proposed framework is not tied to this specific attack and can, in principle, be evaluated under other query-based attack strategies as well.
\end{itemize}

This threat model emphasizes the generality and practicality of our defense framework. Rather than relying on knowledge of a specific classifier or attack formulation, the framework is designed to increase the difficulty of reliably estimating decision boundaries and identifying an optimal evading instance $x^*$.

\section{Proposed Model: RESSAP}
\label{sec:model}
The overall architecture of our proposed model, Robust Ensemble of Selectively Strengthened and Augmented Predictors (RESSAP), is depicted in \textbf{Fig.~\ref{fig:architecture}}. Our framework begins by accepting a pretrained base classifier, denoted as $clf_0$, and a training dataset. This model is assumed to have been developed initially by the data science and/or ML engineering team and is now ready to be assessed for susceptibility to adversarial evasion attacks. The ultimate goals of the framework are twofold: (1) to measure the susceptibility of $clf_0$ to optimal evasion attacks and quantify it based on the number of adversarial queries required to successfully generate an evading instance, and (2) to create a randomized ensemble classifier using the provided training data to increase the adversarial evasion cost, i.e., increasing the number of queries required to craft a successful evading instance.

Multiple weak classifiers are needed to create the ultimate randomized classification system, which is the core promise of this framework. We achieve this by first creating multiple subsets of the feature space and training a weak classifier for each subset (i.e., bag). Then, at runtime, $k$ subsets are selected at random to handle a given classification query. In this context, $clf_0$ is used to compute three feature evaluation metrics based on importance, robustness, and resilience, which guide the construction of $t$ diverse feature subsets.

Next, the training data is vertically split (based on the feature subsets created in the previous step) into multiple bags, with each bag containing the training instances corresponding to a specific feature subset. To improve the generalizability and robustness of the weak classifiers, we then apply noise-based data augmentation within each bag. Finally, a weak classifier is trained on each augmented bag.

This modular structure enables diversity and randomness in both feature representation and model decision-making, thereby making the system significantly harder to bypass through adversarial evasion. In the following subsections, we describe each component of the framework in more detail.

\subsection{Feature Selections and Subset Generations}
The goal of our feature selection process is to construct subsets that are both informative and resilient to adversarial evasion attacks. By identifying features that are significantly discriminatory while remaining stable under adversarial perturbation, we ensure that each classifier in the ensemble is built on a robust and diverse foundation.

In this component, features are selected based on two primary metrics: (1) feature importance, denoted as $FSI$, and (2) feature robustness, denoted as $FSR$. To make the selection process more objective, we introduce a new composite metric, $FSRI$, which integrates both measures to assess the overall resilience of each feature.

\subsubsection{\textbf{Feature Importance} - $FSI$}
This metric measures how crucial each feature is for the base model $clf_0$. Feature importance using feature value permutation is done by measuring the drop in a model's performance when the values of a specific feature are randomly shuffled across all instances in the dataset. This permutation breaks the relationship between that feature and the target variable, allowing us to observe how much the model relied on it to make accurate predictions. A larger drop in performance indicates higher importance of the feature, while little or no change suggests the feature is less relevant.

The normalized feature importance for each feature $j$ is computed as
\begin{equation}
FSI(j) = \frac{\frac{1}{N}\sum_{i=1}^{N}(a - a_{ij})}{\sum_{k=1}^{d}\frac{1}{N}
\sum_{i=1}^{N}(a - a_{ik})},
\end{equation}
where $a$ is the baseline classification accuracy of $clf_0$; $a_{ij}$ is the accuracy after permuting feature $j$ in the $i$-th repetition, $N$ is the number of repetitions, and $d$ is the total number of features.

\subsubsection{\textbf{Feature Robustness} - $FSR$}
This metric evaluates the robustness of individual features based on the pretrained base classifier $clf_0$, by quantifying the change in its performance (e.g., accuracy) when each feature in the training set is perturbed with small random noise. Specifically, we first compute a normalized sensitivity term across features based on the average absolute performance change, and then convert it into a robustness-oriented score by subtracting it from one. This provides a relative robustness measure in which higher values indicate lower sensitivity to perturbation.

\begin{equation}
FSR(j) = 1 - \frac{\frac{1}{N} \sum\limits_{i=1}^{N} \left|a - a_{ij}\right|}{\sum\limits_{k=1}^{d} \frac{1}{N} \sum\limits_{i=1}^{N} \left|a - a_{ik}\right|}
\end{equation}

Equation (2) defines $FSR(j)$, the normalized robustness score of feature $j$, $a$ is the baseline accuracy on the original dataset, $a_{ij}$ is the accuracy after perturbing feature $j$ in the $i$-th repetition using random noise, $N$ is the number of perturbation repetitions, and $d$ is the total number of features. The term is subtracted from one to reflect robustness rather than sensitivity, since the numerator measures how much the model's performance changes when feature $j$ is perturbed.

The only difference between $FSR$ and $FSI$ lies in how the features are altered. Instead of permuting the feature values as in $FSI$, we perturb them by adding small random noise. Additionally, we take the absolute value of the performance change because in robustness analysis we are interested in the magnitude of sensitivity regardless of whether the performance increases or decreases.

\subsubsection{\textbf{Feature Resilience - $FSRI$}}
To fully capture the strength of each feature, we compute a resilience score that combines both its importance and robustness. These metrics are derived entirely by observing how $clf_0$'s performance changes under controlled perturbations of the training instances.

The resilience score for a feature $j$ is then computed by summing its normalized importance and robustness scores:
\begin{equation}
FSRI(j) = FSI(j) + FSR(j)
\end{equation}

A higher $FSRI(j)$ value indicates that feature $j$ is both highly influential to $clf_0$'s decision-making and robust to adversarial or random perturbations. Conversely, features with lower resilience are either less critical, more vulnerable to noise, or both.

\textbf{Subset Generation:}
Once the feature scores are computed using $clf_0$, we dynamically generate feature subsets based on these scores. Each subset $s_l$ is a selected subset of features from the original feature space, and its size $|s_l|$ is determined by:
\begin{equation}
    |s_l| = \left\lceil \frac{l \cdot d}{t} \right\rceil, \quad \text{for} \quad l = 1, 2, \dots, t
\end{equation}
where $d$ is the total number of features, $t$ is the total number of subsets (and classifiers), and $\lceil \cdot \rceil$ denotes the ceiling function to ensure an integer number of features.

Subsets are constructed by selecting the top-ranked features according to three ranking metrics: $FSRI$, $FSI$, and $FSR$. To prioritize the selection of more resilient and informative features while maintaining diversity:
half of the subsets ($\frac{t}{2}$) are constructed using features ranked highest by $FSRI$;
a quarter of the subsets ($\frac{t}{4}$) are constructed using features ranked highest by $FSI$; and
the remaining quarter ($\frac{t}{4}$) are constructed using features ranked highest by $FSR$.

We primarily emphasize resilience (features that are both important and robust) to encourage stronger adversarial resistance, while also including subsets focusing solely on importance (to support clean accuracy) and robustness (to support stability under perturbations).

\subsection{Data Augmentation}
Building upon the feature subsets constructed from the selection metrics, we apply data augmentation to simulate natural variations and minor adversarial perturbations within each bag, encouraging classifiers to learn more stable and generalizable decision boundaries. Data augmentation is applied to each selected feature subset.

Let $\mathbf{X} \in \mathbb{R}^{n \times d}$ denote the original training dataset with $n$ samples and $d$ features. Let $s_l \in S$ denote the set of feature subsets created in the previous step. Let $X_{s_l} \in \mathbb{R}^{n \times |s_l|}$ denote the training dataset sliced based on features in $s_l$. Then data augmentation for each training-set split (i.e., bag) is defined as
\[
\tilde{\mathbf{X}}_{s_l} = \mathbf{X}_{s_l} + \boldsymbol{\epsilon}, \quad \boldsymbol{\epsilon} \sim \mathcal{N}(0, \sigma),
\]
where $\mathcal{N}$ is the Gaussian distribution, and $\sigma$ is the noise level controlling the intensity of the augmentation.

This process mirrors the perturbations used in the computation of $FSR$, ensuring that classifiers trained on these training subsets are exposed to realistic variations and potential adversarial alterations. This step not only simulates natural data variations but also improves the model's resilience by reducing sensitivity to minor adversarial modifications.

\subsection{Classifiers}
Following the data augmentation step, we train an ensemble of $t$ classifiers, each on a distinct, augmented feature subset $\mathcal{S}_l$. Using an ensemble of classifiers trained on diverse subsets enhances model heterogeneity---a critical factor in defending against adversarial attacks. This structure compels adversaries to bypass multiple, independently trained decision boundaries, significantly increasing the complexity of successful evasion. In our proposed model, each classifier $f_l$, where $l \in \{1,\dots,t\}$, is trained on its corresponding augmented subset $s_l$:
\begin{equation}
    f_l = \operatorname{train}\left(\tilde{\mathbf{X}}_{s_l}, \mathbf{y}\right)
    \label{eq:training}
\end{equation}

The ensemble collaboratively strengthens the model’s resilience by incorporating diverse perspectives rooted in feature importance, robustness, and resilience.

\begin{figure}[t]
    \centering
    \includegraphics[width=0.9\linewidth]{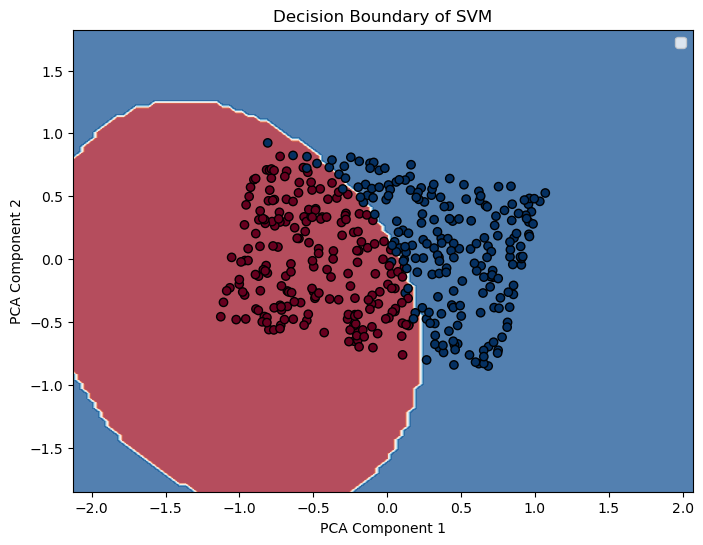}
    \caption{Decision boundary of $clf_0$ on our mock data.}
    \Description{Visualization of the baseline classifier decision boundary on the synthetic mock dataset.}
    \label{fig:svm1}
\end{figure}

\begin{table*}[t]
\centering
\caption{Results on the Mock Dataset (Accuracy \%).}
\label{tab:results}
\begin{tabular}{l|c|c}
\hline
\textbf{Model} & \textbf{Clean Data (Accuracy)} & \textbf{Adversarial Attack (Success Rate)} \\ \hline
SVM & 94.2 & 100 \\
RESSAP-without Feature Selection & 91.7 & 51.45 \\
RESSAP-without Random Classifier Selection & 96.5 & 35.63 \\
RESSAP-without Data Augmentation & \textbf{97.2} & 36.80 \\
RESSAP & 96.5 & \textbf{32.49} \\ \hline
\end{tabular}
\end{table*}

\subsection{Aggregation and Prediction}
At inference time, $k$ classifiers are selected at random to form the prediction set $P$. For each selected classifier $f_l$, the model computes the class probability vector using only the features in $s_l$:
\begin{equation}
    \mathbf{p}_l = f_l\left(\mathbf{x}_{s_l}\right), \quad \text{for } l \in P
    \label{eq:individual_prob}
\end{equation}

The aggregated probability vector is obtained via summation:
\begin{equation}
    \mathbf{p} = \sum_{l \in P} \mathbf{p}_l
    \label{eq:weighted_sum}
\end{equation}
The final predicted class is given by:
\begin{equation}
    \hat{y} = \arg\max_{c} \mathbf{p}(c)
    \label{eq:final_prediction}
\end{equation}
where $\mathbf{p}(c)$ denotes the aggregated probability for class $c$.

\section{Experiments, Results, and Discussion}
\label{sec:experiments}

\subsection{Experimental Setup}
We evaluate our models using a mock dataset consisting of $n=600$ instances and $d=10$ features. The dataset is synthetically generated by first creating a $600 \times 10$ matrix of uniformly distributed random values, which is then scaled by column-specific factors derived from additional uniform random numbers and their tangent transformations. Each feature thus becomes uniformly distributed over a unique range. Binary labels are assigned by computing the $\ell_2$ norm of each sample and thresholding at the median value, ensuring a balanced classification target. The distribution of the dataset is depicted in Fig.~\ref{fig:svm1}. For simplicity, we set the number of classifiers equal to the number of features, i.e., $t=d$, and the number of selected classifiers as $k=t/3$. Exploring the effect of tuning this parameter is left for future work.

To evaluate robustness under adversarial conditions, we apply a query-based evasion strategy using the Multi-Line Search (MLS) method~\cite{nelson2010near}. This algorithm iteratively probes the classifier along weighted feature directions, searching for the minimum-cost perturbation that causes misclassifications. Its efficient binary search mechanism ensures consistent adversarial query generation across all model variants.

We employ two evaluation methods: (1) counting the number of queries required to fool the models, and (2) measuring the success rate of adversarial attacks generated by the MLS attack algorithm. We compare our results against a baseline Support Vector Machine (SVM) classifier as $clf_0$.

\subsection{Results and Discussion}
In our first result, Table~\ref{tab:results} reports (i) the accuracy on clean data and (ii) the success rate of adversarial attacks across different model configurations. As shown, the complete RESSAP model substantially reduces the adversarial attack success rate compared to the baseline SVM (i.e., the initial non-robust $clf_0$), while maintaining high accuracy on clean data. Moreover, the ablation results---in which one key component is removed at a time---highlight the contribution of each mechanism to overall robustness. In particular, omitting feature selection, random classifier selection, or data augmentation consistently leads to weaker robustness. Overall, these results indicate that each component of RESSAP plays an important role in improving resistance to adversarial evasion.

Table~\ref{tab:acc} summarizes robustness from a complementary perspective based on the number of queries required to flip predictions. For each instance, we apply the MLS attack and record the number of queries needed to evade both the baseline SVM and each RESSAP variant. We then compute the percentage of cases in which a given RESSAP variant requires more queries than the SVM. As shown, all RESSAP variants require more queries than the baseline in a majority of cases, and the full RESSAP configuration achieves the highest percentage. This result further supports the claim that RESSAP increases the query cost of successful evasion.

\begin{table}[t]
    \centering
    \caption{Success Rate Based on Number of Queries Compared to the SVM Classifier.}
    \label{tab:acc}
    \begin{tabular}{lc}
        \toprule
        \textbf{Model Variant} & \textbf{Success Rate (\%)} \\
        \midrule
        RESSAP w/o Feature Selection & 61.41 \\
        RESSAP w/o Random Classifier Selection & 65.50 \\
        RESSAP w/o Data Augmentation & 68.26 \\
        RESSAP & \textbf{78.08} \\
        \bottomrule
    \end{tabular}
\end{table}

\begin{figure}[t]
    \centering
    \includegraphics[width=0.92\linewidth]{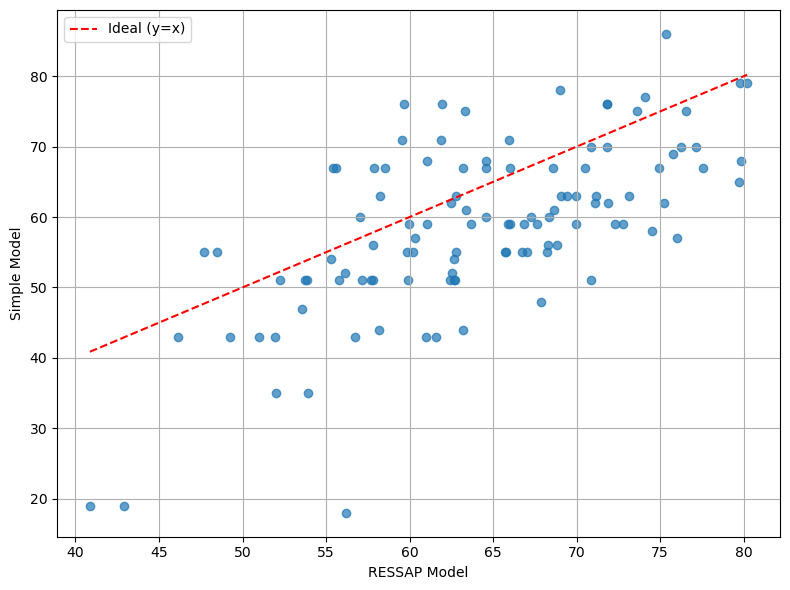}
    \caption{Query Count Comparison: RESSAP vs.\ SVM (Scatter Plot).}
    \Description{Scatter plot comparing per-instance attack query counts for RESSAP (x-axis) versus SVM (y-axis), with an identity line indicating equal query cost.}
    \label{fig:scatter}
\end{figure}

Figure~\ref{fig:scatter} provides a scatter plot comparing the number of queries required per instance to evade RESSAP versus the baseline SVM. Each point corresponds to a single instance; the x-axis shows the number of queries for RESSAP and the y-axis shows the number of queries for SVM. The diagonal identity line $y=x$ indicates equal query cost between the two models. Points below this line correspond to instances for which RESSAP requires more queries than SVM (i.e., higher robustness). As observed, most instances fall below the identity line, indicating that RESSAP is generally more resistant to adversarial perturbations. However, approximately 20\% of instances lie on or above the line, suggesting that for a subset of cases RESSAP offers comparable or slightly lower robustness than the baseline.

\begin{figure}[t]
    \centering
    \includegraphics[width=0.88\linewidth]{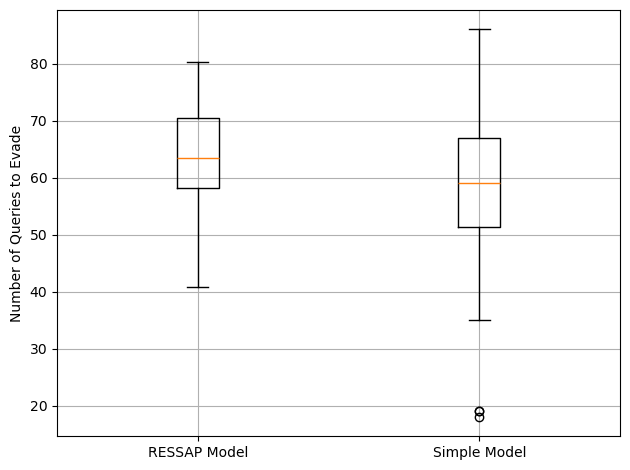}
    \caption{Query Count Comparison: RESSAP vs.\ SVM (Box Plot).}
    \Description{Box plot comparing distributions of attack query counts needed to evade RESSAP and the baseline SVM across instances.}
    \label{fig:box}
\end{figure}

Figure~\ref{fig:box} presents box plots of the query counts required to successfully evade the RESSAP model and the baseline SVM. As shown, the RESSAP model exhibits a higher median query count, indicating that, on average, an adversary must issue more queries to craft a successful evasion. In addition, the interquartile range (IQR) for RESSAP is narrower than that of the baseline, which suggests more consistent robustness across instances. In contrast, the baseline model shows greater variability, including outliers with very low query requirements, highlighting its susceptibility to rapid exploitation. Overall, these results suggest that RESSAP not only increases the difficulty of evasion in most cases, but also reduces variability in the attack cost across different instances.

In summary, our experimental results indicate that combining resilient feature selection, perturbation-based data augmentation, and randomized classifier selection can substantially reduce the success rate of adversarial evasion while maintaining strong clean-data performance. RESSAP leverages an ensemble of classifiers trained on diverse feature subsets and introduces randomness at inference time, making it more difficult for an adversary to reliably estimate decision boundaries and construct effective evading instances.

\section{Conclusion and Future Work}
\label{sec:conclusion}

In this paper, we introduced RESSAP, a robust ensemble framework that combines feature-level selection, data augmentation, and classifier randomization to strengthen classifiers against adversarial evasion attacks. Our experimental evaluation shows that RESSAP improves robustness against adversarial evasion while maintaining strong accuracy on benign inputs.

However, we acknowledge that the current evaluation is limited to a synthetically generated dataset, which may not fully capture the complexity of real-world applications. In addition, we have not yet compared RESSAP against alternative robust architectures that are specifically designed for adversarial settings.

Future work will therefore focus on evaluating the proposed framework on diverse real-world datasets, conducting more extensive comparisons with state-of-the-art robust architectures, and refining the feature selection process to further improve adversarial robustness.

\enlargethispage{2\baselineskip}
\bibliographystyle{ACM-Reference-Format}
\bibliography{main}

\end{document}